\documentclass{article}

\PassOptionsToPackage{numbers, compress}{natbib}
\usepackage[preprint]{neurips_2026}
\makeatletter\renewcommand{\@noticestring}{}\makeatother

\usepackage[utf8]{inputenc}
\usepackage[T1]{fontenc}
\usepackage{url}
\usepackage{microtype}
\usepackage{graphicx}
\usepackage{amsmath}
\usepackage{amssymb}   
\usepackage{booktabs}
\usepackage{subcaption}
\usepackage{hyperref}
\usepackage{xcolor}

\title{RTLScout: Joint Agentic Code and Synthesis Optimization for Efficient Digital Circuits}

\author{%
  Felix Arnold$^{1}$, Ryan Amaudruz$^{1}$, Dimitrios Tsaras$^{2}$, 
  Renzo Andri$^{1}$, Lukas Cavigelli$^{1}$ \\
  $^{1}$Computing Systems Lab, Huawei, Switzerland \\
  $^{2}$Noah's Ark Lab, Huawei, Hong Kong
}

\begin{document}

\newcommand{\AreaPct}{34}
\newcommand{\DelayPct}{38}
\newcommand{\RefineRuns}{50}
\newcommand{\NSweepFront}{14}
\newcommand{\RefineTraj}{700}
\newcommand{\BudgetMin}{20}
\newcommand{\TargetDelays}{800, 1200, 1800}
\newcommand{\TargetDelaysAnd}{800, 1200 and 1800}
\newcommand{\SweepRunsPerDesign}{180}
\newcommand{\ModelName}{GLM 5.2}
\newcommand{\NPhaseOneRuns}{6}
\newcommand{\NPhaseTwoRuns}{6}
\newcommand{\NFresh}{3}
\newcommand{\NSeeded}{3}
\newcommand{\NConcurrent}{2}
\newcommand{\MaxSteps}{30}
\newcommand{\EliteSize}{2}
\newcommand{\StartDelay}{1530}
\newcommand{\CeBestDelay}{1134}
\newcommand{\CampaignMinutes}{136}
\newcommand{\CtxStartK}{36}
\newcommand{\CtxEndK}{69}
\newcommand{\AgentAreaBest}{84}
\newcommand{\NFrontAbc}{5}
\newcommand{\NFrontNoAbc}{5}
\newcommand{\PhasesOneToThreeArea}{81}
\newcommand{\PhasesOneToThreeDelay}{949}
\newcommand{\PFourReduceSentence}{Phase 4 reduces the best area from 81 to 77~\textmu m$^2$ (Fig.~\ref{fig:mockturtle_multirun}).}
\newcommand{\AbcAreaGainPct}{17}
\newcommand{\RRRepsWord}{three}
\newcommand{\RRRepsN}{3}
\newcommand{\RRSteps}{30}
\newcommand{\RRVerilogPct}{10.9}
\newcommand{\RRSpirePct}{25.8}
\newcommand{\RRSpireVsVerilogPct}{16.3}
\newcommand{\GenAreaRange}{12--34}
\newcommand{\GenDelayRange}{10--37}
\newcommand{\GenAdpRange}{22--38}

\newcommand{\emitpaperabstract}{\begin{abstract}\paperabstractbody\end{abstract}}
\newcommand{\paperabstractbody}{%
We present RTLScout, an autonomous system that combines LLM-driven agentic design with logic synthesis optimization and arithmetic architecture selection.
An LLM agent iteratively writes, evaluates, and refines RTL designs, guided by delay and area feedback from Yosys and OpenROAD.
The agent writes Spire, a Python-embedded HDL we introduce, in which optimization intent is expressed locally in the source, selecting logic-synthesis or arithmetic-architecture optimizations per subcircuit.
The four-phase pipeline relies entirely on open-source EDA tools and an open-weights LLM. On an \mbox{IEEE-754}-compliant 16-bit floating-point multiplier with subnormal support, RTLScout reduces area by \AreaPct{}\% and delay by \DelayPct{}\% relative to a starting design and outperforms a commercial-tool reference design on the ASAP7 technology. We show that agentic RTL rewriting and synthesis optimization are complementary, with neither alone reaching the result of the full pipeline.
On 14 RTLRewriter benchmarks, the Spire-based pipeline achieves \RRSpireVsVerilogPct{}\% lower mean per-case Yosys cell count than an otherwise identical Verilog pipeline.
}

\maketitle

\emitpaperabstract

\section{Introduction}

Designing efficient register-transfer level (RTL) hardware requires deep expertise in digital logic, synthesis tool chains, and micro-architectural trade-offs.
Recent advances in large language models (LLMs) have shown that they can generate functionally correct Verilog when guided by natural-language specifications and iterative feedback~\cite{pan_survey}.
However, producing designs that are not merely functionally correct but \emph{Pareto-optimal} across competing physical objectives such as area, delay, and power remains an open challenge.

We present \textbf{RTLScout}, an autonomous design system in which an LLM agent iteratively writes, evaluates, and refines hardware designs through tool use, guided by quantitative feedback from an open-source synthesis and static-timing-analysis (STA) pipeline.
The agent operates within a \emph{multi-run elite pool} framework in which multiple independent runs explore the design space, while the best designs are retained in an elite pool to seed subsequent runs.

\begin{figure}[t]
\vspace{5pt}
    \centering
    \vspace{0.5pt} 
    \includegraphics[width=\columnwidth]{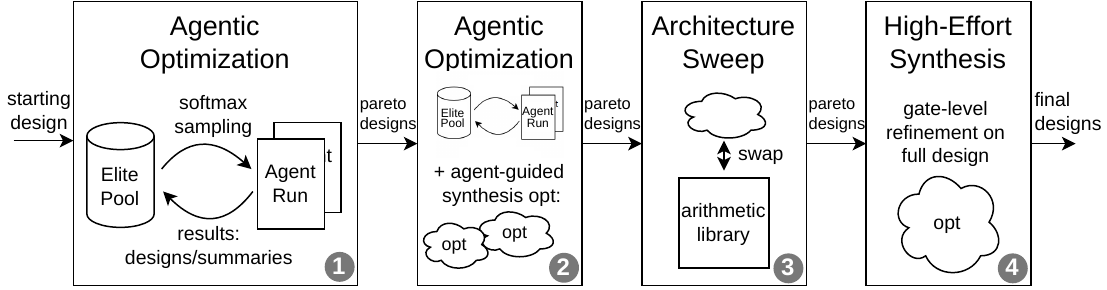}
    \setlength{\abovecaptionskip}{-1pt}
    \caption{End-to-end RTLScout pipeline.
    Phase~3 may alternatively be integrated into Phase~2 through the
    \texttt{@arithmetic\_optimized} decorator, backed by the same arithmetic
    library.
    }
    \label{fig:full_pipeline_flow}
    \vspace{-6pt}
\end{figure}

\begin{figure}[t]
    \centering
    \sbox0{\includegraphics[width=0.49\columnwidth]{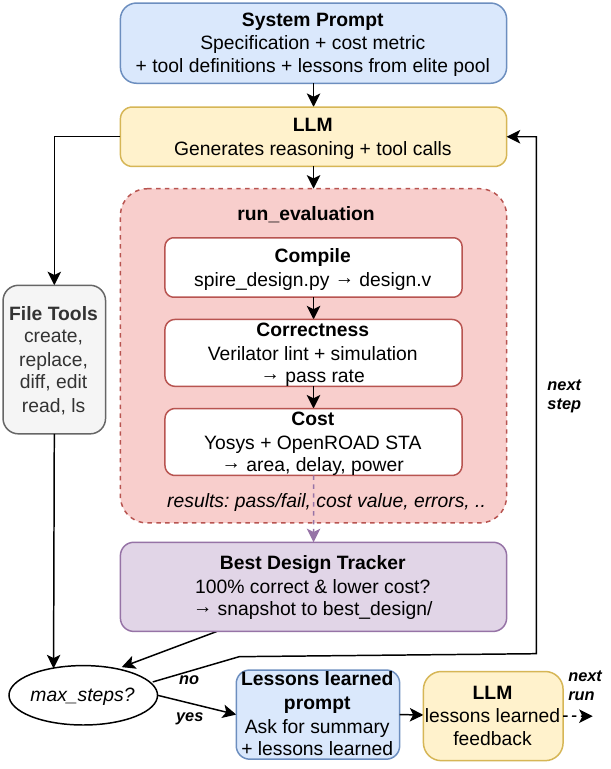}}%
    \begin{subfigure}[b]{0.49\columnwidth}
        \centering
        \begin{minipage}[c][\ht0][c]{\linewidth}
            \centering
            \includegraphics[width=\linewidth]{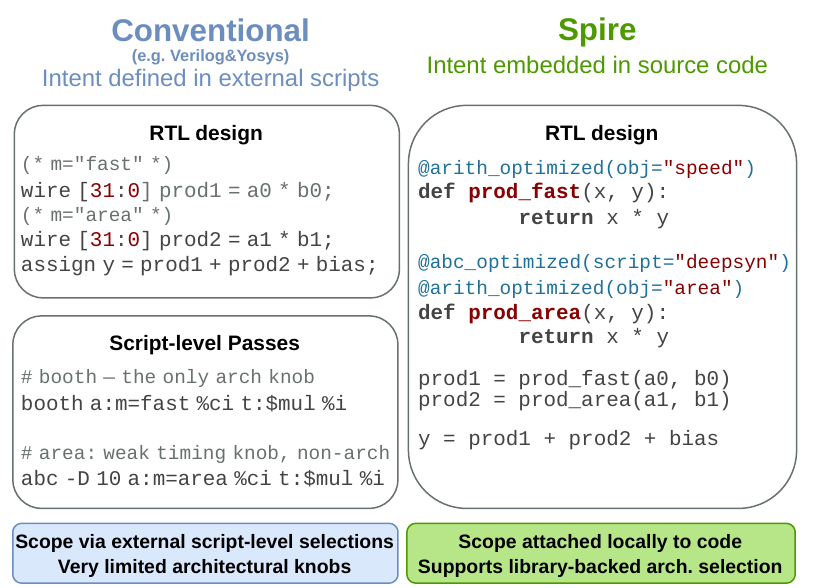}
        \end{minipage}
        \caption{Local annotations make optimization scope explicit.}
        \label{fig:source_level_intent}
    \end{subfigure}\hfill
    \begin{subfigure}[b]{0.49\columnwidth}
        \centering
        \includegraphics[width=\linewidth]{images/agent_flow10d.drawio.pdf}
        \caption{ReAct agent loop.}
        \label{fig:agent_flow}
    \end{subfigure}
    \caption{Agent context and control flow.}
    \label{fig:intent_and_loop}
\end{figure}

A key insight of this work is that multiple phases of optimization are complementary:

\begin{enumerate}
    \item \textbf{Agentic code optimization} rewrites RTL at the algorithmic level---restructuring floating-point logic, eliminating redundant paths, and selecting better micro-architectures.
    \item \textbf{Agent-guided synthesis optimization} extends the agentic code optimization by letting the agent mark subcircuits with a Python decorator that triggers compile-time AIG optimization, reaching locally-optimal Boolean forms that source code changes cannot.
    \item \textbf{Arithmetic architecture sweeps} replace core arithmetic units with building blocks drawn from a library of pre-implemented architectures, sweeping across partial product accumulation strategies, e.g., Wallace and Dadda and prefix adder architectures like Kogge–Stone. 
    \item \textbf{High-effort gate-level refinement}: the Pareto-optimal designs from the architecture sweep are subjected to an additional round of intensive AIG rewriting, yielding further improvements in area and delay.
\end{enumerate}
Fig.~\ref{fig:full_pipeline_flow} illustrates the end-to-end pipeline: the Pareto-optimal designs from each phase seed the next, so later phases build on earlier improvements.
A central enabler of our approach is that optimization choices are represented directly in the source code. 
Fig.~\ref{fig:source_level_intent} illustrates this notion of source-level optimization intent.

This work makes the following contributions:
(1)~\textbf{RTLScout}\footnote{\url{https://github.com/huawei-csl/rtlscout}},
to our knowledge the first autonomous RTL optimization system to unify agentic code rewriting, agent-guided synthesis optimization, and arithmetic architecture search under post-mapping PPA feedback;
(2)~\textbf{source-local optimization intent} as a design paradigm, realized
in \textbf{Spire}\footnote{\url{https://github.com/huawei-csl/spire-hdl}},
a Python-embedded HDL released with this work and built for agentic hardware design, in which optimization choices are expressed locally in the source language;
(3)~a \textbf{multi-run elite pool} framework with z-score softmax seeding and lessons-learned feedback that enables knowledge transfer across independent agent runs; and 
(4)~an \textbf{empirical study} on an IEEE-754 FP16 multiplier and FP32 adder, a posit dot-product unit, and the 14 control-and-datapath cases of the RTLRewriter benchmark, demonstrating competitive area and delay improvements across arithmetic and non-arithmetic designs.

\section{Related Work}

\paragraph{Arithmetic Architecture Design Automation}
Classical approaches to efficient multiplier and adder design include Wallace~\cite{wallace} and Dadda~\cite{dadda} trees for partial product reduction, and prefix adder families such as Kogge--Stone~\cite{kogge_stone}, Brent--Kung~\cite{brent_kung}, and Sklansky~\cite{sklansky}.
Recent work has increasingly automated this design space, including generator- and optimization-based MAC exploration~\cite{easymac,ufo_mac}, reinforcement-learning-based arithmetic tree generation~\cite{arithtreerl,hierarchical_multiplier_rl}, differentiable arithmetic-structure search~\cite{domac,arith_das}, and generative models for prefix-adder and arithmetic-circuit optimization~\cite{prefix_agent,prefix_gpt,ac_refiner}.

\paragraph{LLM-Based RTL Code Generation}

Early work on LLM-based RTL generation targeted functional correctness through domain-specific fine-tuning~\cite{verigen,rtlcoder} and standardized benchmarks of increasing complexity~\cite{verilogeval,rtllm,archxbench}.
More recent multi-agent frameworks decompose complex designs into subtasks with graph-based planning~\cite{mage,verilogcoder} or process unstructured specifications end-to-end through reasoning, progressive coding, and reflection~\cite{spec2rtl_agent}.
Surveys~\cite{pan_survey,yang_survey} cover the field.


\paragraph{PPA-Aware Optimization with LLMs}
PPA-aware LLM systems include OpenROAD-flow tuning agents~\cite{orfs_agent}, LLM-guided e-graph datapath rewriting~\cite{aspen}, iterative PPA-aware Verilog generation~\cite{verippa,drrtl}, and LLM/MCTS-based optimization~\cite{rtlrewriter,delorenzo_mcts}; RTLRewriter~\cite{rtlrewriter} partitions RTL into sub-circuits for cost-aware MCTS rewriting and outperforms Yosys and e-graph optimizers such as E-Syn~\cite{esyn}.
SymRTLO~\cite{symrtlo} adds symbolic reasoning over a rule corpus, and evolutionary approaches~\cite{revolution,hsin2026evolve} co-evolve populations of designs and prompts.
More recent systems refine individual layers of this stack: multi-agent generation loops with evolving memory~\cite{veriagent}, joint correctness/PPA evolution with Pareto selection~\cite{coevo}, retrieval-guided synthesis-flow scheduling~\cite{retrieveschedulereflect}, and test-time reinforcement learning on per-design EDA feedback~\cite{alphartltesttime}.
To our knowledge, none unifies agentic RTL rewriting, synthesis-time Boolean optimization, architecture-level arithmetic exploration, and gate-level refinement end to end.

\section{Method}

\subsection{Agent Architecture}

The RTLScout agent follows the ReAct paradigm~\cite{react}: the LLM generates natural-language reasoning interleaved with structured tool calls, and the system executes those calls and returns results.
The agent has access to file manipulation tools (\texttt{create\_file}, \texttt{replace\_file}, \texttt{apply\_diff}, \texttt{edit\_file}, \texttt{read\_file}, \texttt{ls}), and an evaluation tool (\texttt{run\_evaluation}).
The overall agent loop is illustrated in Fig.~\ref{fig:agent_flow}.

The \texttt{run\_evaluation} tool accepts an optional \texttt{target\_delay} parameter that sets the synthesis and mapping timing constraint, and triggers a multi-stage evaluation pipeline:
(1)~\emph{Compilation}: the Spire Python source (see Section~\ref{sec:mockturtle}) is executed to produce a Verilog netlist;
(2)~\emph{Correctness}: the design is linted with Verilator~\cite{verilator} and simulated against a self-checking testbench;
(3)~\emph{Cost}: PPA metrics are extracted via Yosys synthesis and OpenROAD STA at the specified target delay.
The evaluation result returned to the agent includes: any lint or error messages, simulation pass/fail with the number of failing checks out of the total (e.g., \texttt{Checks (fails/tot): 4185/18874371}), simulation output on failure, the cost value for the target metric, and a full PPA breakdown (area, delay, power), including the worst timing path from STA.
On each successful evaluation, the agent also sees the best cost achieved so far, enabling it to track progress.
A best-design tracker snapshots every evaluation that achieves 100\% correctness at lower cost.

Since the LLM has not been trained on Spire, all knowledge of the domain-specific language (DSL) must be provided through in-context learning via the system prompt.
The system prompt supplies the agent with:
(i)~the design specification, including module name, port widths, and functional requirements;
(ii)~the target cost metric name (e.g., ``area'') embedded directly into the optimization instructions;
(iii)~a complete Spire API reference with code examples, common pitfalls such as width inference, and synthesis hints;
(iv)~a strategy section instructing the agent to first produce a correct design and then iteratively optimize, reverting to the last correct version if a change breaks correctness;
and (v)~the step budget, encouraging the agent to use all available steps for improvement.
When ABC optimization is enabled, additional guidance on the \texttt{@abc\_optimized} decorator is included (see Section~\ref{sec:mockturtle}), including a measured example of its impact.
For seeded runs, the prompt is augmented with lessons-learned summaries from the elite pool (see Section~\ref{sec:elite}).

\subsection{Spire and Gate-Level Optimization}
\label{sec:mockturtle}

\textbf{Spire} is a Python-embedded DSL for hardware design.
Designs are written as Python programs that construct a module graph with width-aware typed signals, and the DSL generates synthesizable Verilog.
This approach suits LLM-driven design: LLMs are trained on large Python corpora, and Python's expressiveness allows loops, functions, and helper modules.

For gate-level logic optimization, we use \textbf{ABC}~\cite{abc}, and in particular its \texttt{\&deepsyn} engine, an anytime portfolio synthesizer that explores randomized optimization scripts and keeps the best intermediate AIG by gate count.
Phase~2 applies curated ABC recipes at compile time to individual subcircuits.
Phase~4 applies \texttt{\&deepsyn} as post-processing to the full candidate designs at substantially higher compute, running \RefineRuns{} randomized \BudgetMin{}-minute trajectories per Pareto-optimal design (see Section~\ref{sec:high_effort_mockturtle}).
We separate Phases~1 and~2 so the agent can first focus on structural and algorithmic RTL rewriting before Phase~2 adds synthesis decorators.

{\emergencystretch=3em
During Phase~2, the system prompt instructs the agent to annotate functions representing logic subcircuits with the \texttt{@abc\_optimized} decorator.
At compile time, decorated functions are converted into an AIG, optimized by ABC, and the resulting netlist is substituted back into the design.
The agent thus operates at two levels simultaneously: restructuring the algorithm at the source level while deciding which subcircuits to delegate to gate-level optimization.
Selective decoration allows assigning \emph{different synthesis objectives and effort to different regions} of a design based on an informed rationale. For instance, the agent can apply an area-minimizing recipe to subcircuits off the reported critical path where timing slack absorbs any local delay increase.
Uniform global synthesis cannot express this per-region trade-off.
The same mechanism is used for arithmetic architecture selection through a separate \texttt{@arithmetic\_optimized} decorator, described in Section~\ref{sec:ari_arch_sweep}.
Fig.~\ref{fig:source_level_intent} illustrates both decorators. ABC-decorated functions cache their results, avoiding repeated optimization of unchanged subcircuits.
\par}

\subsection{Multi-Run Elite Pool Optimizer}
\label{sec:elite}

A single agent run is limited by the LLM's context window, step budget, and path dependence, as early decisions can commit the run to a restricted region of the design space.
To improve exploration, we execute multiple agents within a multi-run elite pool framework, drawing on ideas from evolutionary program search~\cite{funsearch} and verbal reinforcement learning~\cite{reflexion}.

\textbf{Elite pool.}
The pool holds at most $K$ designs ranked by cost.
After each run, the best design is compared with the pool; if it improves upon the worst entry, it replaces it.

\textbf{Seeding strategy.}
Each new run is either \emph{fresh} (initialized from the specification and the provided starting design) or \emph{seeded} (initialized from a design in the pool).
In the seeded case, the pool entry is sampled using a softmax over z-score-normalized costs $z_i = (c_i - \mu)/\sigma$ with selection probability $p_i \propto \exp(-z_i / T)$, where $\mu$ and $\sigma$ are the mean and standard deviation of the pool costs $\{c_i\}$, and $T$ is a temperature parameter. This is similar to the fitness-proportionate selection used in REvolution~\cite{revolution}. Z-score normalization ensures scale invariance across cost metrics.

\textbf{Lessons-learned feedback.}
At the end of each run, a separate LLM call (without tool use) produces a structured summary of what worked, what failed, and lessons for future attempts (Fig.~\ref{fig:agent_flow}).
This summary is stored in the elite pool and injected into the system prompt of all subsequent agents.
Seeded agents receive design files plus lessons, whereas fresh agents receive only lessons and are instructed to try alternative approaches.

\subsection{Arithmetic Architecture Sweep}
\label{sec:ari_arch_sweep}

Phase~3 replaces the core mantissa multiplier and exponent adder with structurally decomposed arithmetic units drawn from a library of classical designs.
Accordingly, in Phases~1 and~2 the agent is instructed to keep multiplication and addition as plain \texttt{*} and \texttt{+} operators, deferring their replacement to Phase~3.
This decouples algorithm-level exploration from arithmetic micro-architecture search, which would otherwise be difficult to exhaustively explore within the agent's step budget.
These structures involve intricate carry propagation and partial product reduction logic, which is error-prone to generate but straightforward to instantiate from a verified library.
The sweep covers (i)~partial-product accumulation using carry-save, Wallace, Dadda, 4:2 compressor, or accumulator trees; (ii)~final-stage adders using Kogge--Stone, Brent--Kung, Sklansky, ripple-carry, or sparse Kogge--Stone (sparsity 2 or 4); and (iii)~area- or speed-optimized full-adder implementations.
Each configuration is evaluated at multiple target delays (\TargetDelays{}~ps), resulting in \SweepRunsPerDesign{} synthesis runs per design ($5\times6\times2\times3$).

\textbf{Decorator-based alternative.}
\label{sec:arith_decorator}
{\emergencystretch=3em
For the generalization experiments in Section~\ref{sec:generalization}, we adopt an alternative formulation that makes Phase~3 obsolete.
Instead of running an arithmetic architecture sweep, the agent uses the Spire \texttt{@arithmetic\_optimized} decorator to access the arithmetic library, parameterized by an objective (\emph{area}, \emph{delay}, or area-delay product \emph{ADP}).
In Phase~2, the agent applies the decorator selectively to functions or operators, deciding both where to invoke it and under which objective.
Phase~1 deliberately stays decorator-free so the agent explores structural rewrites first.
This approach is more flexible than a fixed sweep, better suited to mixed control-and-datapath logic where no single arithmetic operator dominates, and eliminates sweep overhead.
The decorator pairs naturally with the \texttt{@abc\_optimized} decorator (Section~\ref{sec:mockturtle}).
\par}

\section{Experimental Setup}

Our primary benchmark is \textbf{fpmul\_f16}, a fully IEEE-754-compliant 16-bit floating-point multiplier (1 sign, 5 exponent, and 10 fraction bits).
The design supports subnormal inputs and outputs with correct renormalization, implements round-to-nearest-even (IEEE~754 \emph{roundTiesToEven}) including at subnormal boundaries, handles all special cases (zero, $\pm\infty$, NaN propagation, $\infty \times 0 = \text{NaN}$), and produces $\pm\infty$ on overflow.

For the multi-run elite pool experiments that form Phases~1 and~2 of the main pipeline, \ModelName{} serves as the underlying LLM. Separate campaigns with area, delay, and area--delay product (ADP) as cost metric are run, and the resulting Pareto-optimal designs are pooled as input to subsequent phases.
All runs use Spire as the hardware description framework, with the agent writing Python source that the DSL compiles to synthesizable Verilog.

\textbf{Synthesis and STA.}
All synthesis and static timing analysis use the ASAP7 7\,nm predictive FinFET PDK~\cite{asap7} (7.5-track standard-cell library, RVT threshold voltage, TT corner, and NLDM timing models) with Yosys~\cite{yosys} for logic synthesis and OpenROAD~\cite{openroad} for static timing analysis (STA). All reported area and delay values are \emph{post-mapping}, extracted after technology mapping to ASAP7.

\textbf{Functional Verification.}
The starting implementation of \textbf{fpmul\_f16} (Sections~\ref{sec:results} and~\ref{sec:ablation}) is verified exhaustively over all $2^{32}$ input pairs.
Each \texttt{run\_evaluation} call verifies the candidate with a golden-referenced self-checking testbench enumerating $\sim$18.9M directed and structured input pairs (behavioral Verilator simulation of the generated Verilog; the post-mapping netlist is additionally re-simulated within the PPA flow).
\emph{Every} Pareto-optimal fpmul\_f16 design reported in this paper is furthermore verified \emph{exhaustively}: all $2^{32}$ input pairs are simulated at the gate level against a bit-exact golden model.
The input spaces of the \textbf{larger arithmetic designs} (Section~\ref{sec:larger_arith}) exceed $2^{32}$ pairs, making exhaustive verification infeasible. Each candidate is instead checked against a golden reference over at least 1 million directed, structured, and random vectors.
For the \textbf{RTLRewriter benchmarks} (Section~\ref{sec:rtlrewriter}), in-loop testbenches apply $\geq$10\,000 golden-simulated vectors per case.
Every table-backing best design is verified post hoc against its reference using combinational or sequential equivalence checking, and by $10^6$-vector simulation for the two 32-bit multiplier cases.

\section{Results}
\label{sec:results}

\subsection{Multi-Run Optimization}
\label{sec:multi_run}

\begin{figure}[t]
    \centering
    \begin{subfigure}[b]{0.49\columnwidth}
        \centering
        \includegraphics[width=\linewidth]{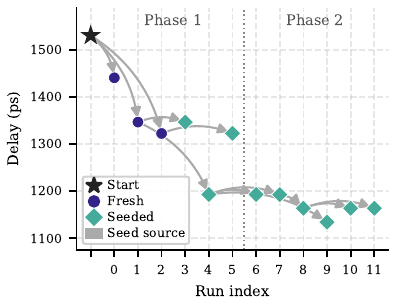}
        \caption{Best delay per run across Phases~1--2; star: starting design, arrows show seeding.}
        \label{fig:cost_evolution}
    \end{subfigure}\hfill
    \begin{subfigure}[b]{0.49\columnwidth}
        \centering
        \includegraphics[width=\linewidth]{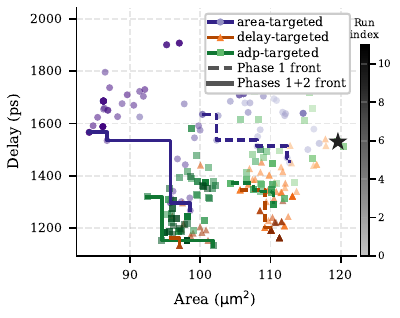}
        \caption{Per-objective results; darker is later run.}
        \label{fig:metric_comparison}
    \end{subfigure}
    \caption{Agent campaign behaviour on fpmul\_f16.}
    \label{fig:agent_campaigns}
\end{figure}

The multi-run elite pool framework enables optimization beyond that achieved by a single run.
Fig.~\ref{fig:cost_evolution} shows the best delay per run across the two delay-targeted campaigns: \NPhaseOneRuns{} Phase~1 runs (\NFresh{} fresh, \NSeeded{} seeded) followed by \NPhaseTwoRuns{} Phase~2 runs, all seeded, with the Phase~2 elite pool initialized from the Phase~1 Pareto front.
Starting from the initial design at \StartDelay{}~ps, the elite pool progressively reduces the best delay to \CeBestDelay{}~ps over \CampaignMinutes{}~minutes of combined campaign time, with lessons-learned feedback enabling knowledge transfer between runs.
Each run begins with a context window of approximately \CtxStartK{}k tokens, dominated by the system prompt containing the Spire API reference, and grows to an average of \CtxEndK{}k tokens by the final evaluation step.

\subsection{Effect of the Cost Metric on the Agent's Pareto Front}
\label{sec:cost_metric}

Fig.~\ref{fig:metric_comparison} overlays the design-space exploration when the agent uses area, delay, or area--delay product (ADP) as the cost metric (agent campaigns only, no architecture sweep), showing the resulting Phase~1 and Phase~1--2 fronts, with the same run configuration per phase (\NPhaseOneRuns{} total runs, \NConcurrent{} concurrent, \MaxSteps{} steps each, elite size \EliteSize{}, temperature 1.0, \NFresh{} fresh-first in Phase~1).
On the Phase~1--2 front, area-targeted runs reduce area to as low as \AgentAreaBest{}\,\textmu m$^2$ at the expense of higher delay, while delay-targeted runs achieve delays as low as \CeBestDelay{}~ps at the expense of larger area.
The per-objective fronts are largely complementary: area-targeted runs dominate the low-area region, while delay-targeted runs dominate the low-delay region. Combining the campaigns therefore broadens the Pareto front.

\subsection{Full Pipeline}
\label{sec:full_pipeline}

\begin{figure}[t]
    \centering
    \begin{subfigure}[b]{0.49\columnwidth}
        \centering
        \includegraphics[width=\linewidth]{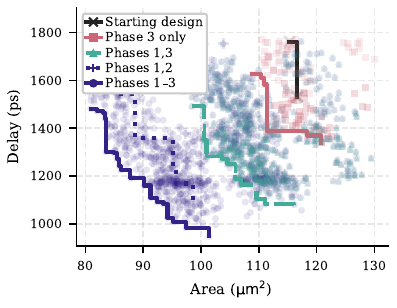}
        \caption{Pareto fronts under Phases~1--3 combinations.}
        \label{fig:full_pipeline}
    \end{subfigure}\hfill
    \begin{subfigure}[b]{0.49\columnwidth}
        \centering
        \includegraphics[width=\linewidth]{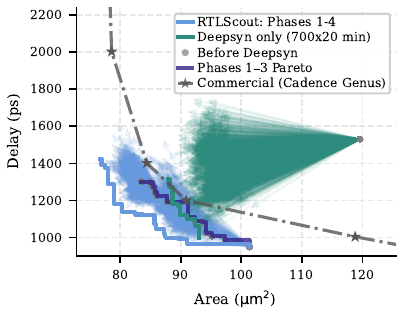}
        \caption{RTLScout: Phases~1--4 vs.\ Deepsyn only.}
        \label{fig:mockturtle_multirun}
    \end{subfigure}
    \caption{Area vs delay across the pipeline on fpmul\_f16. Each point is one synthesis run.}
    \label{fig:pareto_pair}
\end{figure}

Fig.~\ref{fig:full_pipeline} shows the area--delay Pareto fronts for different phase configurations.
The designs fed into the architecture sweep are the Pareto-optimal subset extracted across all agent campaigns---area, delay, and ADP targets, with and without ABC---yielding \NFrontAbc{} Pareto-optimal designs with ABC and \NFrontNoAbc{} without.
Each design is evaluated across all arithmetic configurations and target delays in Phase~3.
Phases~1--3 achieve a best area of \PhasesOneToThreeArea{}~\textmu m$^2$ and a best delay of \PhasesOneToThreeDelay{}~ps, attained by different Pareto-optimal designs.

\textbf{High-effort gate-level refinement (Phase~4).}
\label{sec:high_effort_mockturtle}
In Phase~4, we apply ABC's \texttt{\&deepsyn} with a significantly higher compute budget than the in-agent Phase~2 pass, operating on the full design (as discussed in Section~\ref{sec:mockturtle}).
Each of the \NSweepFront{} Pareto-optimal designs from the architecture sweep is refined by \RefineRuns{} parallel, distinctly seeded \BudgetMin{}-minute Deepsyn trajectories, and the resulting netlists are synthesized at \TargetDelaysAnd{}~ps to explore the area--delay trade-off.
Phase~4 therefore runs Deepsyn for \RefineTraj{}$\times$\BudgetMin{}\,min in total, the same budget as the ``Deepsyn only'' baseline.
\PFourReduceSentence{}

\textbf{Comparison with a commercial-tool reference.}
Fig.~\ref{fig:mockturtle_multirun} includes the IEEE-754-compliant FP16 multiplier implementations from Larsson-Edefors~\cite{larsson2025}, synthesized with Cadence Genus on the same ASAP7 PDK.
Our Phase~4 Pareto front outperforms this commercial-tool reference across the operating range.
Our pipeline uses only open-source EDA tools (Yosys, OpenROAD, Verilator, ABC), whereas~\cite{larsson2025} relies on commercial Cadence Genus and Xcelium.

\section{Ablation Study}
\label{sec:ablation}

Fig.~\ref{fig:full_pipeline} serves as an ablation study, with each curve corresponding to a configuration in which one or more optimization phases are removed.
Table~\ref{tab:ablation} summarizes the extremes achieved for each configuration.

\begin{table}[t]
    \centering
    \caption{Ablation: Extremes across phase combinations on fpmul\_f16. \emph{agent}: Phase~1 RTL rewriting; \emph{synth}: agent-guided synthesis via decorators; \emph{arith}: arithmetic architecture library. Values are the best over the sweep target delays (800/1200/1800\,ps). In parentheses: in-loop agent values (Section~\ref{sec:cost_metric}).}
    \label{tab:ablation}
    \begin{tabular}{@{}lcccccc@{}}
        \toprule
        Configuration & \multicolumn{4}{c}{Phase} & Best area (\textmu m$^2$) & Best delay (ps) \\
        \cmidrule(lr){2-5}
         & 1 & 2 & 3 & 4 & & \\
        \midrule
        Starting design &  &  &  &  & 116 & 1530 \\
        Phases~1,2 (agent+synth) & \checkmark & \checkmark &  &  & 86 (84) & 1073 (1134) \\
        Phases~1,3 (agent+arith) & \checkmark &  & \checkmark &  & 98 & 1083 \\
        Phase~3 only (arith only) &  &  & \checkmark &  & 109 & 1339 \\
        Phases~1--3 (agent+synth+arith) & \checkmark & \checkmark & \checkmark &  & 81 & 949 \\
        Deepsyn only (700$\times$20\,min) &  &  &  & \checkmark & 89 & 1000 \\
        Deepsyn only (700$\times$40\,min) &  &  &  & 2$\times$ & 88 & 1000 \\
        \midrule
        RTLScout: Phases~1--4 & \checkmark & \checkmark & \checkmark & \checkmark & \textbf{77} & \textbf{949} \\
        \bottomrule
    \end{tabular}
    \vspace{-10pt}
\end{table}

\textbf{Agent removal.}
The starting design using only the arithmetic sweep (``Phase~3 only'' in Table~\ref{tab:ablation}) performs substantially worse than any agent-optimized variant, indicating that the algorithmic improvements arise from the LLM-driven optimization rather than architecture search alone.

\textbf{ABC-agent removal.}
Without agentic gate-level optimization, the agent designs with arithmetic sweep (``Phases~1,3'' in Table~\ref{tab:ablation}) trail the Phases~1--3 results. The ABC phase provides a \AbcAreaGainPct{}\% area reduction beyond what is achieved by the Phase~1 agent and architecture sweep alone.

\textbf{Arithmetic sweep removal.}
Agent designs with ABC but without the architecture sweep (``Phases~1,2'' in Table~\ref{tab:ablation}) fall short of the Phases~1--3 results. The arithmetic sweep further improves the extremes by replacing the \texttt{*}/\texttt{+} operators with structurally decomposed arithmetic units.

\textbf{High-effort synthesis only.}
To evaluate whether Phase~4 alone could substitute for the preceding phases, we apply it directly to the unmodified starting-point design at matched compute (``Deepsyn only'' in Table~\ref{tab:ablation}).
The standalone result forms a narrow cluster---a substantial reduction from the starting point, but far less diverse and unable to match the agent-optimized configurations.

None of the ablated configurations approaches the full pipeline.
Thus, each phase contributes improvements that are inaccessible to the others, demonstrating the complementary nature of the optimization phases.


Table~\ref{tab:budget} reports the per-phase compute and API budget with Deepsyn-only baseline listed separately.

\begin{table}[t]
    \centering
    \caption{Per-phase compute and API budget for fpmul\_f16 (method rows; comparison baselines listed below the total and excluded from it). Compute priced at \$0.04 per core-hour.}
    \label{tab:budget}
\begin{tabular}{@{}lrrrr@{}}
\toprule
Phase & API (\$) & Compute (core-h) & Compute (\$) & Total (\$) \\
\midrule
Phase 1 (agent) & 25.67 & 3.4 & 0.13 & 25.81 \\
Phase 2 (agent+synth) & 26.88 & 7.1 & 0.29 & 27.17 \\
Phase 3 (arch sweep) & 0.00 & 49.1 & 1.97 & 1.97 \\
Phase 4 (Deepsyn refine) & 0.00 & 243.0 & 9.72 & 9.72 \\
Verification & 0.00 & 10.3 & 0.41 & 0.41 \\
\midrule
Total (method) & 52.55 & 312.9 & 12.52 & 65.07 \\
\midrule
Baseline: Deepsyn only & 0.00 & 242.4 & 9.70 & 9.70 \\
\bottomrule
\end{tabular}

\end{table}

\section{Generalization}
\label{sec:generalization}

We evaluate generalization along two axes: larger arithmetic designs (Table~\ref{tab:generalization-circuits}) and the RTLRewriter control-and-datapath suite (Table~\ref{tab:best-cells}).
In these experiments, Phase~3's arithmetic sweep is integrated into the \texttt{@arithmetic\_optimized} decorator (Section~\ref{sec:arith_decorator}) rather than run separately.

\subsection{Larger Arithmetic Designs}
\label{sec:larger_arith}

\begin{table}[t]
\centering
\caption{Generalization to other circuits: best per-phase area (\textmu m$^2$), delay (ps) and area-delay product ($10^3$\,\textmu m$^2\cdot$ps) for each benchmark, all Spire, evaluated at a common set of target delays. Each cell is an independent best over that phase's evaluated designs; \textbf{bold} marks the best value per column.}
\label{tab:generalization-circuits}
\resizebox{\columnwidth}{!}{%
\begin{tabular}{@{}lccccccccc@{}}
\toprule
Phase & \multicolumn{3}{c}{fpadd-32} & \multicolumn{3}{c}{PDPU-16 (comb.)} & \multicolumn{3}{c}{PDPU-16 (pipelined)} \\
\cmidrule(lr){2-4}\cmidrule(lr){5-7}\cmidrule(lr){8-10}
 & Area & Delay & ADP & Area & Delay & ADP & Area & Delay & ADP \\
\midrule
Starting design & 145.0 & 1980 & 294.4 & 708.4 & 5307 & 3759.5 & 1044.2 & 865 & 907.4 \\
Phase 1 (agent) & 122.8 & 1550 & 208.2 & 674.6 & 3402 & 2469.7 & 736.0 & 811 & 779.4 \\
+Phase 2 (agent+synth+arith) & 108.5 & \textbf{1527} & 203.2 & 633.0 & 3362 & 2429.1 & 702.3 & 778 & 725.3 \\
RTLScout: +Phase 4 (Deepsyn refine) & \textbf{100.9} & \textbf{1527} & \textbf{192.7} & \textbf{624.4} & \textbf{3351} & \textbf{2339.0} & \textbf{690.4} & \textbf{775} & \textbf{707.9} \\
\midrule
Deepsyn only, equal compute & 102.7 & 1762 & 198.1 & 654.0 & 5518 & 3653.2 & 989.9 & 796 & 808.9 \\
\bottomrule
\end{tabular}
}
\end{table}

\paragraph{Floating-point adder.}
\label{sec:fpadd}
To evaluate generalization beyond multiplication, we target
\textbf{fpadd\_f32}, a fully IEEE-754-compliant binary32 adder (8-bit exponent,
23 fraction bits) with subnormal, rounding, and special-case support,
roughly 20\% larger in starting-design area than the main study's 16-bit multiplier.
Its input space rules out exhaustive verification, so each
candidate is checked against a golden reference over 16M directed and
structured input pairs.

\paragraph{Posit dot-product unit.}
\label{sec:pdpu}
The posit dot-product-accumulate unit (\textbf{PDPU}~\cite{pdpu}) (16 lanes,
posit(8,2) elements with a posit(16,2) accumulator) is taken from its reference
RTL implementation and ported to Spire. The required behavior is bit-exactness with respect to
that reference, and each candidate is checked against it over 2M directed and
random vectors. We evaluate both the combinational top and the 5-stage pipelined
top ($5.9\times$ and $8.7\times$ the main study's 16-bit multiplier).

\paragraph{Results.}
Table~\ref{tab:generalization-circuits} reports per-phase best area, delay, and
area-delay product for all three designs, evaluated at a common set of target
delays. Every phase contributes incremental improvements on every benchmark, consistent with the multiplier: the full pipeline reduces area by \GenAreaRange{}\% and delay by
\GenDelayRange{}\%.
The Deepsyn-only baseline on the starting design at matched compute
yields consistently worse results. Combining agent rewriting with high-effort
synthesis again achieves superior results.

\subsection{RTLRewriter Benchmarks}
\label{sec:rtlrewriter}


\begin{table*}[t]
\centering
\caption{Best per-phase Yosys cell count on the 14 RTLRewriter cases. \textbf{Base}: shipped baseline; \textbf{RTLR}: paper target; \textbf{P1} is decorator-free structural exploration, \textbf{P2} enables \texttt{@arithmetic\_optimized}/\texttt{@abc\_optimized} and seeds from P1. $\Delta_{1\!\to\!2}$ within-language P1$\to$P2; $\Delta_\text{vs R}$ vs.\ RTLR; $\Delta_\text{S/V}$ Spire P2 vs.\ Verilog P2 (cross-language, same pipeline). Negative $=$ reduction; \textbf{bold} $=$ strict row minimum, \underline{underline} $=$ tied for minimum. Agent cells additionally show mean$\pm$std of the per-repetition bests across $n{=}3$ repetitions; the leading number is the best repetition. $\Sigma$: column totals; $\Delta\Sigma$: $\Delta$ between the totals; $\mu\Delta$: mean of the per-case $\Delta$. All percentages are computed from the means of the per-repetition bests.}
\label{tab:best-cells}
\resizebox{\textwidth}{!}{%
\begin{tabular}{@{}l@{}r@{\hspace{5pt}}r@{\hspace{5pt}}r@{\hspace{5pt}}r@{\hspace{5pt}}r@{\hspace{5pt}}r@{\hspace{5pt}}r@{\hspace{5pt}}r@{\hspace{5pt}}r@{\hspace{5pt}}r@{\hspace{5pt}}r@{\hspace{5pt}}r@{}}
\toprule
 &  & \multicolumn{5}{c}{\textbf{Ours (Verilog)}} & \multicolumn{5}{c}{\textbf{Ours (Spire)}} &  \\
\cmidrule(lr){3-7} \cmidrule(lr){8-12}
\# & RTLR & Base & P1 & P2 & $\Delta_{1\!\to\!2}$ & $\Delta_\text{vs R}$ & Base & P1 & P2 & $\Delta_{1\!\to\!2}$ & $\Delta_\text{vs R}$ & $\Delta_\text{S/V}$ \\
\midrule
1 & 14 & 18 & \underline{10} {\scriptsize$10{\pm}0$} & \underline{10} {\scriptsize$10{\pm}0$} & +0.0\% & -28.6\% & 18 & \underline{10} {\scriptsize$13{\pm}5$} & \underline{10} {\scriptsize$10{\pm}0$} & -21.1\% & -28.6\% & +0.0\% \\
2 & 11475 & 11824 & 11272 {\scriptsize$11.3k{\pm}12$} & 11262 {\scriptsize$11.3k{\pm}18$} & -0.1\% & -1.8\% & 11899 & 11469 {\scriptsize$11.5k{\pm}6$} & \textbf{8499} {\scriptsize$8.63k{\pm}166$} & -24.8\% & -24.8\% & -23.5\% \\
3 & 974 & 1220 & 661 {\scriptsize$662{\pm}1$} & 653 {\scriptsize$659{\pm}5$} & -0.5\% & -32.4\% & 1220 & 662 {\scriptsize$662{\pm}0$} & \textbf{426} {\scriptsize$426{\pm}1$} & -35.6\% & -56.2\% & -35.3\% \\
4 & 1213 & 1462 & 823 {\scriptsize$827{\pm}3$} & 823 {\scriptsize$826{\pm}3$} & -0.1\% & -31.9\% & 1461 & 829 {\scriptsize$829{\pm}0$} & \textbf{556} {\scriptsize$571{\pm}14$} & -31.2\% & -53.0\% & -30.9\% \\
5 & 49 & 49 & 32 {\scriptsize$35{\pm}3$} & \textbf{23} {\scriptsize$32{\pm}8$} & -8.5\% & -34.0\% & 49 & 37 {\scriptsize$37{\pm}0$} & 37 {\scriptsize$37{\pm}0$} & +0.0\% & -24.5\% & +14.4\% \\
6 & 129 & 129 & 115 {\scriptsize$124{\pm}8$} & 115 {\scriptsize$119{\pm}8$} & -4.0\% & -7.5\% & 129 & 129 {\scriptsize$129{\pm}0$} & \textbf{114} {\scriptsize$114{\pm}1$} & -11.4\% & -11.4\% & -4.2\% \\
7 & 354 & 403 & 263 {\scriptsize$283{\pm}34$} & 255 {\scriptsize$261{\pm}5$} & -7.9\% & -26.3\% & 351 & 265 {\scriptsize$288{\pm}27$} & \textbf{240} {\scriptsize$256{\pm}14$} & -11.2\% & -27.8\% & -2.0\% \\
8 & 370 & 370 & 370 {\scriptsize$370{\pm}0$} & 370 {\scriptsize$370{\pm}0$} & +0.0\% & +0.0\% & 370 & 343 {\scriptsize$361{\pm}16$} & \textbf{337} {\scriptsize$341{\pm}3$} & -5.6\% & -7.9\% & -7.9\% \\
9 & 32 & 40 & 40 {\scriptsize$40{\pm}0$} & 39 {\scriptsize$40{\pm}1$} & -0.8\% & +24.0\% & 55 & 21 {\scriptsize$40{\pm}17$} & \textbf{15} {\scriptsize$26{\pm}15$} & -33.6\% & -17.7\% & -33.6\% \\
10 & 42 & 56 & 37 {\scriptsize$40{\pm}4$} & 36 {\scriptsize$37{\pm}1$} & -7.6\% & -12.7\% & 30 & 7 {\scriptsize$7{\pm}1$} & \textbf{6} {\scriptsize$7{\pm}1$} & -9.1\% & -84.1\% & -81.8\% \\
11 & \underline{24} & \underline{24} & \underline{24} {\scriptsize$24{\pm}0$} & \underline{24} {\scriptsize$24{\pm}0$} & +0.0\% & +0.0\% & 33 & \underline{24} {\scriptsize$24{\pm}0$} & \underline{24} {\scriptsize$24{\pm}0$} & +0.0\% & +0.0\% & +0.0\% \\
12 & 14674 & 14960 & 14454 {\scriptsize$14.5k{\pm}38$} & 14454 {\scriptsize$14.5k{\pm}22$} & -0.1\% & -1.3\% & 15001 & 14536 {\scriptsize$14.5k{\pm}1$} & \textbf{10861} {\scriptsize$11.0k{\pm}211$} & -24.0\% & -24.8\% & -23.7\% \\
13 & \underline{1} & \underline{1} & \underline{1} {\scriptsize$1{\pm}0$} & \underline{1} {\scriptsize$1{\pm}0$} & +0.0\% & +0.0\% & \underline{1} & \underline{1} {\scriptsize$1{\pm}0$} & \underline{1} {\scriptsize$1{\pm}0$} & +0.0\% & +0.0\% & +0.0\% \\
14 & \underline{2} & 3 & \underline{2} {\scriptsize$2{\pm}0$} & \underline{2} {\scriptsize$2{\pm}0$} & +0.0\% & +0.0\% & 3 & \underline{2} {\scriptsize$2{\pm}0$} & \underline{2} {\scriptsize$2{\pm}0$} & +0.0\% & +0.0\% & +0.0\% \\
\midrule
\textbf{$\Sigma$} & 29353 & 30559 & 28104 & 28067 &  &  & 30620 & 28335 & \textbf{21128} &  &  &  \\
\textbf{$\Delta\Sigma$} &  &  & -7.7\%$^{\dagger}$ & -7.9\%$^{\dagger}$ & -0.3\% & -4.2\% &  & -7.0\%$^{\dagger}$ & -29.7\%$^{\dagger}$ & -24.4\% & -26.8\% & -23.6\% \\
\textbf{$\mu\Delta$} &  &  & -18.9\%$^{\dagger}$ & -20.5\%$^{\dagger}$ & -2.1\% & -10.9\% &  & -21.5\%$^{\dagger}$ & -32.8\%$^{\dagger}$ & -14.8\% & -25.8\% & -16.3\% \\
\bottomrule
\end{tabular}%
}
\par\smallskip\raggedright\footnotesize $^{\dagger}$\,$\Delta$ relative to the Verilog Base for both languages (numbers compare directly between languages).
\end{table*}

We utilize the 14 short-bench cases from RTLRewriter~\cite{rtlrewriter}, which include chain adders, ALU/FSM control, and mux-tree redundancy.
Each case includes an unoptimized Verilog baseline and a reference RTLRewriter (RTLR) implementation.
{\emergencystretch=3em
The Phase~3 sweep is folded into the Phase~2 \texttt{@arithmetic\_optimized} decorator and Phase~4 is omitted here, so the pipeline reduces to Phases 1–2. Phase~1 performs decorator-free structural exploration, and Phase~2 enables the arithmetic and ABC decorators.
We use \RRRepsWord{} independent single-run repetitions per phase (\RRSteps{} steps each) across two HDL variants (Verilog and Spire), reporting Yosys post-\texttt{synth} cell counts.
\par}

Both pipelines outperform the RTLRewriter reference designs (Table~\ref{tab:best-cells}), achieving mean per-case reductions in Yosys cell count of $\RRVerilogPct{}\%$ for Verilog and $\RRSpirePct{}\%$ for Spire (means of per-repetition bests, $n=\RRRepsN{}$). Spire additionally outperforms the Verilog agent by a mean of $\RRSpireVsVerilogPct{}\%$ per case.
This result contrasts with PyHDL-Eval~\cite{pyhdleval}, which found lower LLM pass rates for Python-embedded DSLs than Verilog on 168 specification-to-RTL problems. Our controlled comparison for iterative RTL optimization instead finds parity without decorators and an advantage for Spire once they are included (P1/P2 $\Delta$ rows in Table~\ref{tab:best-cells}).
A separate campaign that optimizes the Yosys gate-level transistor estimate reproduces the pattern (Appendix~\ref{app:rr_transistors}, Table~\ref{tab:best-transistors}).

\section{Conclusion}

We presented RTLScout, an autonomous system that combines agentic RTL rewriting with logic synthesis-level optimization. The framework implements both a sequential combination, where high-effort synthesis refines the agent’s designs as a separate phase, and a coupled form, where the agent marks subcircuits for synthesis optimization and selects recipes through local decorators. 
Similarly, an arithmetic-architecture library is used either in an exhaustive sweep or via a more flexible decorator that lets the agent choose its target and objective. Both source-level decorators are enabled by Spire, a Python-embedded HDL we introduced in this work.

The ablation study shows that agentic RTL rewriting and synthesis optimization are complementary. Another axis of diversity we uncovered is the objective given to the agent: area, delay, and area-delay product (ADP) each lead to different Pareto-optimal designs and pooling these broadens the Pareto front.
Overall, the full pipeline consistently reduces area and delay across four IP blocks by 10\% to 40\%, outperforms a commercial-tool reference on the FP16 multiplier, and reduces cell count on the RTLRewriter cases by a mean of \RRSpirePct{}\% relative to the published reference. Future work may be to make the decorators more effective by estimating the impact of subcircuits on the overall design, guiding where to apply the decorators and which recipe to choose.

\clearpage

\bibliographystyle{unsrtnat}
\bibliography{references}


\clearpage
\appendix
\section{Additional Campaign Data}

\subsection{Generalization benchmarks: per-phase fronts}
\label{app:gen_fronts}

\begin{figure}[h]
    \centering
    \includegraphics[width=\textwidth]{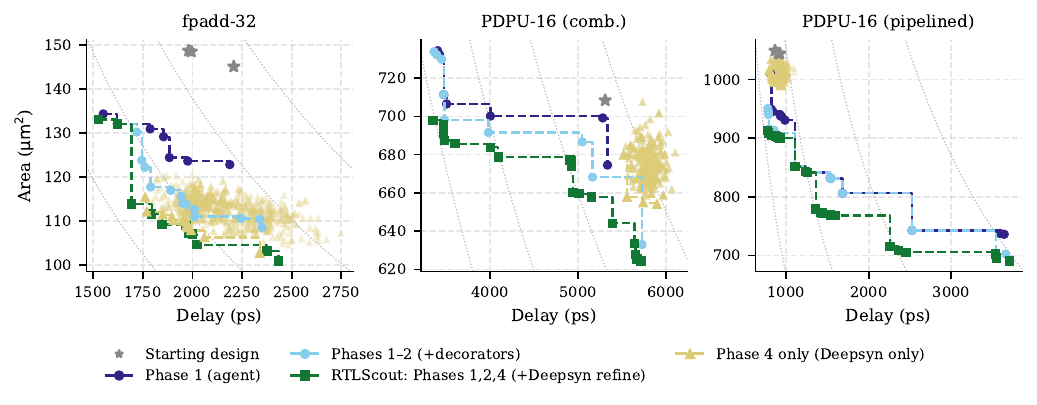}
    \caption{Per-phase area--delay fronts for the generalization benchmarks
    (Spire, evaluated at the reporting target delays of
    Table~\ref{tab:generalization-circuits}). Staircases are each phase's Pareto front; dotted gray curves are iso-ADP contours.}
    \label{fig:gen_fronts}
\end{figure}

Fig.~\ref{fig:gen_fronts} shows the per-phase area--delay fronts behind
Table~\ref{tab:generalization-circuits}. Each successive phase advances the
front on every benchmark, while Phase~4 only (Deepsyn only) remains well behind
the full pipeline on all three designs, yielding both a weaker Pareto front and
a narrower range of area--delay trade-offs.

%
%

\subsection{Transistor count on the RTLRewriter cases}
\label{app:rr_transistors}

To test whether the result of Section~\ref{sec:rtlrewriter} is
specific to the cell-count metric, we ran a new campaign with the optimization
target set to the Yosys gate-level transistor estimate, with an otherwise
identical setup (both languages, three repetitions, 30 steps).
Table~\ref{tab:best-transistors} reports the results. Since the RTLRewriter
paper publishes no transistor targets, $\Delta_\text{vs B}$ compares against
each language's own baseline. The pattern of Table~\ref{tab:best-cells}
reproduces: near parity at Phase~1 ($\Delta\Sigma$ of $-7.7\%$ vs.\ $-7.8\%$
relative to the Verilog base) and a Spire advantage once the decorators are
enabled ($\Delta_{1\to2}$ of $-0.1\%$ vs.\ $-19.3\%$), leaving Spire's
Phase-2 total $19.3\%$ below Verilog's.

\noindent\textbf{Case index.} Case numbers in Tables~\ref{tab:best-cells} and~\ref{tab:best-transistors} denote the RTLRewriter modules 1:~\texttt{add3}, 2:~\texttt{commutativity\_subexpression}, 3:~\texttt{multi\_constant\_multiplication}, 4:~\texttt{multi\_constant\_multiplication2}, 5:~\texttt{adder\_bit\_width}, 6:~\texttt{adder\_subexpression}, 7:~\texttt{alu\_subexpression}, 8:~\texttt{multiplier\_bitwidth}, 9:~\texttt{example1}, 10:~\texttt{example3}, 11:~\texttt{mux\_dead\_code}, 12:~\texttt{communtativity\_subpexpression2}, 13:~\texttt{mux\_type3}, 14:~\texttt{mux\_type4}.


\begin{table*}[t]
\centering
\caption{Best per-phase Yosys transistor count on the 14 RTLRewriter cases. Counts are Yosys's post-\texttt{synth} CMOS transistor estimate (\texttt{stat -tech cmos}). \textbf{Base}: shipped baseline; \textbf{P1} is decorator-free structural exploration, \textbf{P2} enables \texttt{@arithmetic\_optimized}/\texttt{@abc\_optimized} and seeds from P1. $\Delta_{1\!\to\!2}$ within-language P1$\to$P2; $\Delta_\text{vs B}$ final best vs.\ that language's own \textbf{Base}; $\Delta_\text{S/V}$ Spire P2 vs.\ Verilog P2 (cross-language, same pipeline). The RTLRewriter paper reports no transistor target, so the RTLR column is omitted and $\Delta_\text{vs B}$ takes the place of the cell table's $\Delta_\text{vs R}$. Negative $=$ reduction; \textbf{bold} $=$ strict row minimum, \underline{underline} $=$ tied for minimum. Agent cells additionally show mean$\pm$std of the per-repetition bests across $n{=}3$ repetitions; the leading number is the best repetition. $\Sigma$: column totals; $\Delta\Sigma$: $\Delta$ between the totals; $\mu\Delta$: mean of the per-case $\Delta$. All percentages are computed from the means of the per-repetition bests.}
\label{tab:best-transistors}
\resizebox{\textwidth}{!}{%
\begin{tabular}{@{}l@{}r@{\hspace{5pt}}r@{\hspace{5pt}}r@{\hspace{5pt}}r@{\hspace{5pt}}r@{\hspace{5pt}}r@{\hspace{5pt}}r@{\hspace{5pt}}r@{\hspace{5pt}}r@{\hspace{5pt}}r@{\hspace{5pt}}r@{}}
\toprule
 & \multicolumn{5}{c}{\textbf{Ours (Verilog)}} & \multicolumn{5}{c}{\textbf{Ours (Spire)}} &  \\
\cmidrule(lr){2-6} \cmidrule(lr){7-11}
\# & Base & P1 & P2 & $\Delta_{1\!\to\!2}$ & $\Delta_\text{vs B}$ & Base & P1 & P2 & $\Delta_{1\!\to\!2}$ & $\Delta_\text{vs B}$ & $\Delta_\text{S/V}$ \\
\midrule
1 & 256 & 128 {\scriptsize$128{\pm}0$} & 128 & +0.0\% & -50.0\% & 256 & \underline{96} {\scriptsize$96{\pm}0$} & \underline{96} {\scriptsize$96{\pm}0$} & +0.0\% & -62.5\% & -25.0\% \\
2 & 78258 & 75418 {\scriptsize$75.5k{\pm}111$} & 75418 {\scriptsize$75.4k{\pm}51$} & -0.0\% & -3.6\% & 79774 & 75506 {\scriptsize$75.5k{\pm}70$} & \textbf{60330} {\scriptsize$60.5k{\pm}155$} & -19.9\% & -24.1\% & -19.8\% \\
3 & 7872 & 3192 {\scriptsize$3.85k{\pm}1.15k$} & 3192 {\scriptsize$3.85k{\pm}1.15k$} & +0.0\% & -51.0\% & 7960 & 3192 {\scriptsize$3.19k{\pm}0$} & \textbf{3092} {\scriptsize$3.09k{\pm}0$} & -3.1\% & -61.2\% & -19.8\% \\
4 & 8742 & 4334 {\scriptsize$4.35k{\pm}14$} & 4326 {\scriptsize$4.34k{\pm}11$} & -0.3\% & -50.4\% & 8724 & 4344 {\scriptsize$4.35k{\pm}10$} & \textbf{4168} {\scriptsize$4.28k{\pm}99$} & -1.5\% & -50.9\% & -1.2\% \\
5 & 268 & 258 {\scriptsize$261{\pm}6$} & \textbf{256} {\scriptsize$257{\pm}1$} & -1.5\% & -4.0\% & 268 & 268 {\scriptsize$268{\pm}0$} & 262 {\scriptsize$262{\pm}0$} & -2.2\% & -2.2\% & +1.8\% \\
6 & 882 & 846 {\scriptsize$864{\pm}25$} & 844 {\scriptsize$863{\pm}27$} & -0.1\% & -2.2\% & 882 & 882 {\scriptsize$882{\pm}0$} & \textbf{828} {\scriptsize$833{\pm}8$} & -5.6\% & -5.6\% & -3.5\% \\
7 & 2304 & \underline{1684} {\scriptsize$1.83k{\pm}125$} & \underline{1684} {\scriptsize$1.81k{\pm}117$} & -0.8\% & -21.4\% & 2314 & 1898 {\scriptsize$2.11k{\pm}255$} & 1836 {\scriptsize$2.03k{\pm}278$} & -4.0\% & -12.4\% & +11.8\% \\
8 & 2538 & 2468 {\scriptsize$2.51k{\pm}37$} & 2462 {\scriptsize$2.51k{\pm}40$} & -0.1\% & -1.2\% & 2542 & 2538 {\scriptsize$2.54k{\pm}2$} & \textbf{2392} {\scriptsize$2.40k{\pm}14$} & -5.5\% & -5.5\% & -4.3\% \\
9 & 160 & 78 {\scriptsize$125{\pm}42$} & \underline{74} {\scriptsize$123{\pm}43$} & -2.1\% & -23.3\% & 280 & 78 {\scriptsize$183{\pm}91$} & \underline{74} {\scriptsize$97{\pm}37$} & -46.9\% & -65.2\% & -20.7\% \\
10 & 110 & 54 {\scriptsize$64{\pm}10$} & 54 {\scriptsize$61{\pm}8$} & -5.2\% & -44.8\% & 220 & \underline{26} {\scriptsize$34{\pm}7$} & \underline{26} {\scriptsize$30{\pm}7$} & -11.8\% & -86.4\% & -50.5\% \\
11 & 192 & \underline{160} {\scriptsize$160{\pm}0$} & \underline{160} {\scriptsize$160{\pm}0$} & +0.0\% & -16.7\% & \underline{160} & \underline{160} {\scriptsize$160{\pm}0$} & \underline{160} {\scriptsize$160{\pm}0$} & +0.0\% & +0.0\% & +0.0\% \\
12 & 101572 & 97774 {\scriptsize$97.9k{\pm}85$} & 97720 {\scriptsize$97.7k{\pm}28$} & -0.1\% & -3.8\% & 101614 & 97902 {\scriptsize$97.9k{\pm}0$} & \textbf{77248} {\scriptsize$77.3k{\pm}16$} & -21.1\% & -24.0\% & -21.0\% \\
13 & 12 & \underline{6} {\scriptsize$6{\pm}0$} & \underline{6} & +0.0\% & -50.0\% & \underline{6} & \underline{6} {\scriptsize$6{\pm}0$} & \underline{6} {\scriptsize$6{\pm}0$} & +0.0\% & +0.0\% & +0.0\% \\
14 & 36 & \underline{24} {\scriptsize$24{\pm}0$} & \underline{24} & +0.0\% & -33.3\% & 36 & \underline{24} {\scriptsize$24{\pm}0$} & \underline{24} {\scriptsize$24{\pm}0$} & +0.0\% & -33.3\% & +0.0\% \\
\midrule
\textbf{$\Sigma$} & 203202 & 186424 & 186348 &  &  & 205036 & 186920 & \textbf{150542} &  &  &  \\
\textbf{$\Delta\Sigma$} &  & -7.7\%$^{\dagger}$ & -7.8\%$^{\dagger}$ & -0.1\% & -7.8\% &  & -7.8\%$^{\dagger}$ & -25.7\%$^{\dagger}$ & -19.3\% & -26.3\% & -19.3\% \\
\textbf{$\mu\Delta$} &  & -24.9\%$^{\dagger}$ & -25.4\%$^{\dagger}$ & -0.7\% & -25.4\% &  & -24.4\%$^{\dagger}$ & -32.7\%$^{\dagger}$ & -8.7\% & -31.0\% & -10.9\% \\
\bottomrule
\end{tabular}%
}
\par\smallskip\raggedright\footnotesize $^{\dagger}$\,$\Delta$ relative to the Verilog Base for both languages (numbers compare directly between languages).
\end{table*}

\end{document}